\documentclass[fleqn,twoside]{article}
\usepackage{amsmath}
\usepackage{espcrc2}

\DeclareMathOperator*{\fd}{\rightarrow}

\title{
 A  phenomenological $\pi ^-$p scattering length from pionic hydrogen }
 
\author{ T.E.O.~Ericson $^{a}$\thanks{Corresponding author.  \mbox{\it E-mail }:
torleif.ericson@cern.ch}, B. Loiseau$^{b}$ and S. Wycech$^{c}$\\
\addressmark{$^a$ Theory Division, Physics Department, CERN, CH-1211 Geneva 23, Switzerland, and 
 TSL, Box 533, S-75121 Uppsala, Sweden, \\
$^b$ Laboratoire de Physique Nucl\'eaire et de Hautes 
\'Energies\thanks{Unit\'e de Recherche des Universit\'es
Paris 6 et Paris 7, associ\'ee au CNRS}, Groupe Th\'eorie,\\
\hspace{0.25cm}Univ. P. \& M. Curie, 4 Pl. Jussieu, F-75252 Paris, France \\
$^c$ Soltan Institute for Nuclear Studies,
 PL-00681 Warszawa,
 Poland}}
        
\begin{document}
 \begin{abstract}
We derive a closed, model  independent,  expression for the electromagnetic  correction factor
 to  a  phenomenological hadronic scattering length  $a^h$ 
 extracted from  a hydrogenic atom. It is obtained  in  a non-relativistic approach and 
 in the limit of a short ranged hadronic  interaction to terms of order 
$\alpha ^2\log \alpha $ using an
 extended charge distribution.  A hadronic $\pi $N 
scattering length  $a^h_{\pi ^-p}=0.0870\,(5)m_{\pi }^{-1}$ is deduced leading to a $\pi $NN
 coupling constant from the GMO relation 
$g_c^2/(4\pi )=14.04\,(17)$.

 \end{abstract}

\maketitle
\section {Introduction }
The strong interaction energy shifts $\epsilon _{1s}$ and total decay width $\Gamma_{1s} $
in pionic hydrogen  have been measured to a remarkable precision~\cite{SCH01}
\begin{eqnarray}
\epsilon _{1s} = 
  [-7.108\!\pm\! 0.013({\rm stat})\!\pm\! 0.034({\rm syst})]~{\rm eV},
\label {epsilonexp}\\
\Gamma _{1s}  = 
 [0.868 \pm 0.040 {\rm (stat)}\pm 0.038{\rm (syst)}]~{\rm eV}.
\label {gammaexp}
\end{eqnarray}
 It is well known~\cite{DESER54,TRUMAN61}  that the (complex) strong interaction shift  in the 1s state of hadronic atoms is closely linked to the 
(complex) elastic threshold scattering  amplitude $a$ defined 
in the absence of the Coulomb field. We refer in the following to this quantity  as the (complex) scattering length.
This is conventionally expressed in the ratio of the shift to the Bohr energy $E _B=-m\alpha ^2/2$:
\begin{eqnarray}
\frac { \epsilon _{1s}}{E_B}= \frac { \epsilon ^0_{1s}}{E_B}(1+\delta _{1s} )=4m~\alpha ~a~(1+\delta _{1s}),
\label{Truemanformula}
\end{eqnarray}
where  $\delta _{1s}$ conveniently measures the deviation of the shift from the lowest order estimate  
\begin{equation}
\epsilon _{1s}^0=-\frac {4\pi }{2m} \ \phi ^2_B(0)\ a. \label{lowestorder}
\end{equation}
 
 Here $\phi_B(r)$ is the non-relativistic 1s  Bohr wave function  of a point charge and 
 $m$  the reduced mass, which in the present case is that of the $\pi^-$p system. 
It is important to understand the  correction  $\delta _{1s}$  transparently and  reliably  to an accuracy matching the high experimental precision, since the hadronic $\pi $N scattering lengths are key testing quantities for chiral physics.
In addition, they  are needed phenomenologically to about 1 \%  for the precision determination of the $\pi $NN coupling constant using the GMO relation~\cite{ERICSON02}. 
 
The standard conversion of experimental data to a scattering length uses the potential approach of Sigg et al.~\cite{SIGG96}, which describes the $\pi $N interaction in terms of coupled equations using physical pion masses and an isospin invariant non-diagonal potential matched to scattering lengths calculated  by setting the neutral pion mass equal to the charged one.
This gives $\delta _{1s}({\rm Sigg})=(-2.1\pm 0.5) \%$. 
The procedure is model dependent  and it is not consistent with the $\pi $N low-energy expansion~\cite{ERICSON02}. 
Their results must therefore be used with caution. 

The classical way to obtain Eq. (\ref{Truemanformula})  is based on  analytical approaches using Coulomb wave functions (see Refs.~\cite{DESER54,TRUMAN61,LAM69,CAR92,HOL99} and references therein). 
To our knowledge  these papers do not  explore the effect of  the  extended  charge distribution
 on the strong interaction shift. 
This paper discusses this question.

 The authors of Refs.~\cite{LYUBO00,GASSER02} have calculated the 
ground state energy of the $\pi^-p$ system in 
the framework of QCD+QED, using effective field theory  (EFT)
techniques. The shift of the ground  state energy  
is related  to the scattering lengths in pure QCD, 
evaluated in the isospin symmetric limit $m_u=m_d$ 
The corresponding  
correction $\delta_\epsilon$ in the energy shift
is evaluated
in the framework of Chiral Perturbation Theory (ChPT). At leading 
order~\cite{LYUBO00}, one has  $\delta_\epsilon$
=$(-4.3\pm2.8)\%$, while the next-to-leading order result~\cite{GASSER02}
 is $\delta_\epsilon$= $(-7.2\pm 2.9)\%$.
The uncertainty in $\delta_\epsilon$  is due to  
 the poor knowledge 
of one of the low-energy constants occurring in 
the effective theory. The 
correction $\delta_\epsilon$ is also 
considerably larger than what was found in  Ref.~\cite{SIGG96}. 
This EFT approach makes inherently no distinction between the
atomic corrections due to Coulomb effects discussed here and other contributions.  We come back in 
section 4 to a comparison  with the present work.
 
Our aim here concerns only the connection of  the strong atomic energy shift to  the scattering length 
$a^h$ defined as  the one  which  would be observed if the Coulomb field of the extended charge could be removed and 
considered as due to an external source. This scattering length is directly related to the one appearing 
in forward dispersion relations for
 $\pi $N scattering.  In this spirit, no correction is made for the internal e.~m. 
 contributions to the masses.   
The physical scattering lengths for the $\pi ^{\pm }$n scattering correspond closely to  the present 
 definition, neglecting  the very small e.~m. correction from the   charged pion interaction with the  neutron
 charge distribution. 
These scattering lengths   with physical masses are  natural 
'observables'  for the study of  isospin breaking. 
The (complex) $\pi ^+$n scattering length coincides  with the $\pi ^-$p one in the limit of exact charge symmetry: 
it has the corresponding open  charge exchange channel  $\pi ^+{\rm n}\rightarrow \pi ^0 {\rm p}$ and the 
corresponding  open radiative decay channel $\pi ^+{\rm n}\rightarrow \gamma {\rm p}$.
This definition is different from  that of the QCD scattering length  used in  the EFT approach. 
We use here the $r$-representation, which is more transparent for the present problem than the equivalent 
momentum representation. 
Since  the $\pi ^-$p atom is highly non-relativistic, most of the discussion will be made using 
non-relativistic concepts.
The result will be expressed in terms of the   empirical on-shell parameters of the 
$\pi $N  low-energy expansion. 
For the electromagnetic corrections, this approach gives  intuitively  interpretable  
expressions,  exact up to terms in
 $\alpha ^2 \log \alpha $,  provided the Coulomb potential of the extended 
 charge varies little   over the range of the strong interaction.

In section~2  we solve the problem exactly to all orders in $\alpha $ in the limit of a short ranged strong 
interaction  with the charge located to a spherical shell. 
  The correction 
 for an arbitrary charge distribution is then derived 
perturbatively
to the same order in $\alpha $ as in the EFT expansion~\cite{GASSER02}.
Corrections for the finite interaction range are explored. 
We explicitly include the correction for the vacuum polarization. 
In section~3 we discuss the magnitudes of the corrections and their physical structure.  
In section~4, we  compare our results to those of previous approaches.

We denote by $E _{fs}, E$ and $\epsilon _{1s}=E-E_{fs}$  the 1s finite size e.m. binding energy, the total 1s binding energy  with strong interaction and finite size, and the strong interaction shift, respectively. The  non-relativistic wave numbers are 
$\kappa _B,~ \kappa _{fs}$ and $\kappa $ for $E_B, E_{fs}$ and $E$, respectively. 
Since this paper concerns atomic corrections we also use the Bohr radius  $r_B=\kappa _B^{-1}=(m\alpha )^{-1}$.
\section{A Model for the $\pi ^-$p Atom} 
The aim of this section is to explore the consequences of the
extended charge distribution in a pedagogically transparent and soluble model of the  $\pi ^-$p system.  
This will serve as a prototype for the later  more general discussion and  
it will reveal the nature of the contributions to  the correction term in Eq.~(\ref {Truemanformula}). 
In the absence of the Coulomb potential the threshold expansion for the angular
momentum $l=0$, typical of  a weak scattering length, is related to the phase
shift $\delta _{l=0}$ and to the momentum $q$ by the relation
\begin{equation}
\frac {\tan \delta  ^h_{l=0}}{q}=a^h+b^hq^2+....
\label{rangeexp}
\end{equation}
Here $a^h$ is the hadronic scattering length and $b^h$ is the range parameter. 

The model is constructed as follows. 
We first consider  the case of a single channel. 
This avoids the complications of  several  open channels  with different masses 
 for which the equivalent of the  single channel  scattering length is ambiguous.
The generalization will be discussed later.  
The charge is assumed to be  concentrated to a spherical shell of radius $R$, outside the range of the 
hadronic interaction.
The system is taken to be non-relativistic.
For the moment we neglect the effect of the vacuum polarization potential.
 As defined,  this problem can be solved  exactly in terms of the 
 on-shell hadronic s-wave scattering amplitude, although we 
 will only evaluate it  for contributions to the  correction term up to order 
$\alpha ^2\log \alpha $. In this case the definition of the scattering length is clear.
 The model does not have problems with the intermingling of the Coulomb and hadronic interaction, contrary to other descriptions. Such problems are 
particularly  acute in any description with a pointlike charge 
distribution, since the Coulomb 
interaction is then divergent  at $r=0$. 

Inside a typical shell radius R of the order of 1 fm, the Coulomb potential is constant with $V_C(R)=-\alpha/R\simeq 1.4 $ MeV.  This motivates the non-relativistic approximation in this region to order $\alpha $.  
 The inside   wave number $q_c$ is constant:
\begin{equation}
 q_c^2=\frac {2m\alpha }{R} -\kappa ^2\simeq \frac {2m\alpha }{R}.
\end{equation}
The $1s$ binding energy   $E~\simeq -3.2$~keV  is negligible compared to the Coulomb field and the strong interaction inside  the charge distribution region, although its exact value governs the scale of the atom. 
The external 1s wave function for $r\geq R$ is  a Whittaker function ${\cal  W}_{\lambda ;1/2} (z)\label {GRA}$ (see e. g.,~\cite {GRA}, Eq.~(9.237))  with 
$\lambda =\kappa _B/\kappa $ and $z=2\kappa r$. 
The inside wave function for $r\leq R$  is a standing wave outside the strong interaction region.  
Neglecting terms of order$(1-\lambda )^2\simeq (\alpha ma^h)^2\simeq 10^{-6}$,  one has for $r\leq  R$,
\begin{equation}
u_{in}(r)=N \left[ \frac {\sin \ (q_cr)}{q_c}+\frac {\tan \delta ^h_{l=0}}{q_c}\cos\  (q_cr)\right] \label{u_i}
\end{equation}
and for $r\geq R$, 
$$
u_{out}(r)  =  (4\pi)^{1/2})^{-1} \kappa \exp (-z/2)
$$
\begin{equation}
\label{u_o}
 \times\!\left\{z\big[(1\!+(1-\lambda )(1-\gamma -\log~z)\big]\!+\frac {1-\lambda
}{\lambda }\right\}\!.
\end{equation}
Here $\gamma =0.577...$  is the Euler constant. 
Note that the term  $\tan (\delta ^h_{l=0})$ is  determined by the `physical' hadronic  phase shift  in the absence of the Coulomb field taken  at the energy $-V_C(R)$.
The wave function corresponding to Eqs.~(\ref {u_i}) and (\ref{u_o}) is normalized to order 
$\alpha ^2$. 

The energy shift $ \epsilon _{1s} $ produced by the strong interaction is obtained by matching the logarithmic derivative of the wave function at the radius $R$.  
In accordance with standard practice, it is  defined as the difference between the total binding energy $E $ and the electromagnetic binding energy  with a finite size charge distribution~\cite {SIGG96}. 
This corresponds  to the removal of the 'scattering length' corresponding to the extended charge distribution: 
\begin{equation}
a_{fs}=-\frac {\alpha mR^2}{3};~\epsilon _{1s}^{fs}=\frac {2\pi \alpha R^2}{3}\phi ^2_B(0). 
\label {R2}
\end{equation}
 Expanding the exact analytical expression to terms of order 
$\alpha ^{2}¥\log \alpha $ by a straightforward  algebraic 
calculation gives the following  correction factors, where the hadronic scattering length $ a^h$ 
takes the place of $a$ in Eq.~(\ref{Truemanformula}):
\begin{eqnarray}
\delta _{1s} & =  & 
 -2\frac {R}{r_B} 
  +2\frac {a^h}{r_B}\left[2-\gamma -\log (2\alpha mR)\right]\nonumber\\
& + &\frac {2m\alpha }{R} \frac {b^h}{a^h}. \label{deltapoint2}
\end{eqnarray}
 This expression serves as a guide for the later generalizations. 

The assumption of a zero range hadronic interaction is unnecessary in our simple model. 
An interaction of any shape will give the same result provided its range is smaller than R.
This follows from the matching condition for the wave functions (\ref{u_i}) and (\ref{u_o}), 
which is  only required at $R$, such that any interaction with the same hadronic  scattering near-threshold amplitude, $a^h+q_c^2b^h$, gives the same result.

The  terms in Eq.~(\ref {deltapoint2}) have  a clear physical interpretation.
The extended charge wave function at $r=0$ in the absence of strong interactions is
 $\phi _{in}(0)= \phi _B(0)(1-R/r_B+..)$ to the present order in $\alpha$.
It is a better starting approximation than the wave function of the Bohr atom in  
Eq.~(\ref {lowestorder}), which then should be multiplied by a factor $\left(1-R/r_B+..\right)^2\simeq 1-2R/r_B+..$. 
This accounts for the first term in Eq.~(\ref{deltapoint2}).   
The second term proportional to $a^h$ is a renormalization due to the external wave function which is changed at $R$ by the hadronic scattering itself 
by a factor $1+2a^hm\alpha [2-\gamma -\log (2\alpha mR)]$. 
The outside wave function is determined by the energy shift. 
The matching of the inside and outside wave functions of Eqs.~(\ref{u_i}) and  (\ref{u_o}) gives  near the origin 
\begin{eqnarray}
u_{in}(r) & = & \{1+2a^hm\alpha [2-\gamma -\log (2\alpha mR)]\}\nonumber \\
& \times & (r+a^h)\ \phi _{in}(0). 
 \end{eqnarray}
This result agrees with that obtained by matching the logarithmic derivative at $R$. 
This  factor    has little sensitivity to the exact value of the radius~R. 
The leading $\alpha \log \alpha  $  part of the term in $a^h$ is well known from previous approaches and has also been found in the EFT approach where it corresponds to a "loop" term~\cite{LYUBO00,GASSER02}.

 The last term in Eq.~(\ref {deltapoint2})  follows  from gauge invariance with the replacement $E \rightarrow E -V_C(0)$ in the scattering amplitude~\cite{ERICSON02}. Alternatively, and more intuitively,  it follows using  the correct energy at the point of interaction. 
This  is not   the binding energy, but  the finite depth of the Coulomb potential of the extended charge. (For the corresponding effect in  higher Z pionic atoms, see Refs.~\cite{Ericson1982}-\cite{FRI03}).

Exactly the same reasoning as for the 1s state can be applied to hadronic energy shift 
$\epsilon _{ns}$ in states of any n. The correction factor $\delta _{ns}$ is defined in complete analogy to 
Eq.~(\ref {Truemanformula}):
\begin{equation}
\frac {\epsilon _{ns}}{E _{ns}}= \frac {\epsilon _{ns}^0}{E_{ns}}(1+\delta _{ns})=
4\frac { a^h}{nr_B}(1+\delta _{ns}),
\label {defdeltans}
\end{equation}
where $E _{ns}=-m\alpha ^2/(2n^2)$ and the convenient comparison shift $\epsilon _{ns}^0$ is the simplest perturbative expression for the energy shift
\begin{equation}
\epsilon _{ns}^0= -\frac {4\pi }{2m} \ \phi ^2_{B;ns}(0)\ a^h=\frac {\epsilon _{1s}^0}{n^3}.
\end{equation}
One has to order $\alpha ^2\log \alpha $ in the correction
\begin{equation}
\delta _{1s}-\delta _{ns}=2\frac{a^h}{r_B}\left(1-\frac {2}{n}+\sum _1^{n}\frac {1}{k}-\log~n\right).
\label {deltan-s}
\end{equation}
In this expression all the dependence on the parameter $R$ has disappeared, which reflects that all the short-ranged physics is identical but for a renormalization factor.
In the limit  $n\rightarrow \infty $
\begin{equation}
\delta _{1s}-\delta _{\infty s}=2\frac{a^h}{r_B}(1+\gamma ),
\label {delta1-infinity}
\end{equation}
where the correction term is given by
\begin{eqnarray}
\delta _{\infty s } & =  & -2\frac {R}{r_B}
+  2\frac{a^h}{r_B}\left[1-2\gamma -\log~\left(\frac{2R}{r_B}\right)\right] \nonumber\\
& + &\frac {2m\alpha }{R}\frac {b^h}{a^h}.\label {deltainfinity}
\end{eqnarray}

This semi-classical limit for  $\kappa _n\equiv (\kappa _B/n) \rightarrow ~0$  corresponds 
to the Coulomb scattering length~\cite {PRESTON} $a^c=a^h(1+\delta_{\infty s})$ in our model.
For the present case of a $\pi ^-$p atom the numerical difference in the correction terms for different values of $n$ in Eqs.~(\ref  {deltan-s}) and (\ref {delta1-infinity}) is less than
$10^{-3}$ and of little practical importance.

\noindent {\bf  Arbitrary Charge Distribution}\\
The result  (\ref{deltapoint2}) is the prototype for more general  charge distributions.  
The difference between the Coulomb potential $V_{Cfs}$  for a charge 
 distribution from the observed $\pi ^-$ and proton form factors, $\rho (r)$, and $V_{CR} $ corresponding to that for the spherical shell of radius $R$, $\rho _R(r)$, gives  a perturbative potential, which includes the f.s. charge density associated  with the anomalous magnetic moment:
$$
\delta V_C(r)  = V_{Cfs}(r)-V_{CR}(r)
$$
\begin{equation}
\hspace*{1cm}=  -  \alpha \int\limits_{r}^{\infty }\left(\displaystyle\frac {1}{r'}-\displaystyle\frac {1}{r}\right)\delta \rho (r')4\pi r^{'2}dr',  \label{Coulpot}\end{equation}
where $\delta \rho (r')=\rho (r)-\rho _R(r')$.  Applying this 
perturbation to our soluble model gives a net 
correction independent of $R$ to the present order in $\alpha $. The 
explicit  calculation leads to 
the following four changes in our model results.
First, 
 the  e.~m. finite size energy shift~(\ref {R2}) is changed with the substitution of the model $R^2$ by
 $\left<r^2\right>_{em}=\left<r_p^2\right>_{em}+\left<r_{\pi }^2\right>_{em}=1.15(2)$ fm$^2$ as in Ref.~\cite{SIGG96}.  
Likewise, the value of the overall Coulomb potential at the origin changes from the model value 
$\alpha /R$ to $\alpha \left<1/r\right>_{em}$ and the wave function squared at the origin changes 
it value from $(1-2\alpha R +..)$ to  $(1-2\alpha \left<r\right>_{em}+..)$.
 Finally, the term  $\log R$  is replaced by 
$\left<\log ~ r\right>_{em}$.
The changes are independent of the hadronic interaction.  

In the case of a single channel and in the hadronic zero range limit, the corrections are:
$$
 \delta _{1s}  = -2\frac {\left< r\right>_{em}}{r_B} + 2\frac{a^h}{r_B}\!\left(2\!-\!\gamma\! -\left<\!\log \frac {2r}{r_B}\!\right>_{em}\right)
$$
\begin{equation}
\hspace*{0.9cm}+2m\alpha \left<\!\frac {1 }{r}\!\right> _{em}\frac {b^h}{a^h}
\equiv  \delta^{\left< r\right>}+ \delta^{c} + \delta^{g}.\label{deltapoint1}
\end{equation}
We now introduce the correction $\delta ^{vp}$ for the vacuum polarization~\cite {EIR00}. 
The first order vacuum polarization is described by a potential proportional to $\alpha ^2$
with a range much larger than that of the hadronic and charge distribution ones and it is insensitive to the strong interaction dynamics. 
The joint extended  Coulomb potential and  vacuum polarization one is  a perfectly justifiable alternative to the point Coulomb potential  as the starting point for the wave function in  Eq.~(\ref {Truemanformula}).  
The  square of the  unperturbed wave function at the origin changes  by 
$\delta ^{vp}=2\delta \phi _{vac}(0)/\phi _B(0)=0.48\%~$ due to  vacuum polarization~\cite{EIR00} and by $-$0.85 \% from the extended charge (see Table~\ref {table:1}). 
This result is model-independent and it agrees with the prior numerical value implicit in 
Ref.~\cite {SIGG96}. 
 
In the derivation of the correction factors,  we nowhere used that these quantities should be real. 
We can therefore take the energy shift to be complex, 
$\epsilon _{1s}  -  {\rm i}\Gamma _{1s}/2 $, with a hadronic complex scattering length 
$a^h_r+{\rm i}a^h_i$. 
The complex energy shift is related to the corresponding correction factors by $a(1+\delta)\rightarrow a_r(1+\delta _r)+{\rm i}a_i(1+\delta _i)$, as in Eq.~(\ref{Truemanformula}). 
The imaginary part  corresponds to absorptive phenomena.  
In the notation of Eq.~(\ref {deltapoint1}):
$$
\delta_{1s,i}= -2\frac {\left< {r}\right>_{em}}{r_B} 
+4\frac{a_r^h}{r_B}\left(2-\gamma -\left<\log \frac {2r}{r_B}\right>_{em}\right)
$$
\begin{equation}
\hspace*{0.8cm}+\  2m\alpha \left<\frac {1}{r}\right>_{em}\frac {b^{h}_i}{a^{h}_i}\equiv  \delta^{\left< r\right>}+ 2\delta^{c} + \delta^{g}_i.
\label{deltai}
\end{equation}
Here the imaginary amplitudes $a^{h}_i$ and $b^{h}_i$ refer to  any absorptive channel such as the $\pi ^-$p charge exchange scattering.
Note the additional factor 2 in the middle term as compared to that for the real case  in Eq.~(\ref {deltapoint1}).  
Since $a_i<<|a_r|$, the change in $\delta _{1s,r}$  due to absorption is negligible.

We conclude that most of the corrections to the width are due to the change of the wave function at origin: it is  important to use wave functions corresponding to the finite size and vacuum polarization potentials. 
In addition, the non-linear renormalization term  must also be included, but only the real part of the scattering length is relevant. 
To these  should be added the amplitude change due to the gauge term in analogy to the case  for the energy shift.

\noindent {\bf Coupled channels}\\
The  $\pi^- p$ atom is  a coupled system of   the continuum $\pi^o n$ and $\gamma n$  channels in addition to the $\pi ^-$p one. 
These  three  channels  are denoted by  indices $ i~(j) = c,o,f$, respectively. 
The  low-energy expansion  in multiple channel systems  is defined in terms of  energy dependent (symmetric) K-matrices which enter the  standing wave solutions. 
The formalism is described briefly below and  it is illustrated for the 2-channel situation.  
The single channel  becomes a special case. 
The standing waves at  distances  larger  than the charge radius $ r> R$  are   defined as~\cite{COR62}
\begin{equation} 
\label{d1} 
u^i_j    =  u_r^i \delta_{i,j}  + K^c_{i,j}  u_s^j,   
\end{equation} 
where  
$ K^c_{i,j} $  are the  "Coulomb-corrected"  K-matrix elements. 
The wave functions   $u^c_r$ and $u^c_s $ are defined in terms of   the standard regular  and singular Coulomb functions F and G, respectively (see  e.g. Refs.~\cite {LAM69,COR62}). 
In the limit  $\alpha \rightarrow 0$ and $q$ fixed, these solutions correspond to $\sin (qr)/q$ and $\cos (qr)/q$, respectively.  Furthermore
\begin{equation}
\label{b1} 
u^c_r   =  \frac {F(r)}{C(\eta)q}
 \fd\limits_{r\rightarrow \infty}   \frac{\sin(\varphi)}{ C(\eta)q };
\end{equation}
 \begin{eqnarray}
 u^c_s   =   G(r)C(\eta) -2 \eta h(\eta) \frac{ F(r)}{ C(\eta ) }\qquad\qquad  \nonumberÊ&&\\
 \fd\limits_{r\rightarrow \infty}   C (\eta)
\cos(\varphi)-\frac {2\eta h(\eta) \sin(\varphi)}{ C(\eta )},&&
 \end{eqnarray}
where $ \eta = z z' \alpha m /q $  and  $ q \eta =-\kappa _B $.
In these equations the digamma function $ \psi(x)$ defines  $h(\eta)= [\psi(i\eta)+ \psi(-i\eta)]/2 - \log( \eta^2) /2 $ and $ C^2(\eta )= 2\pi \eta/ [\exp (2\pi \eta )-1] $ is the standard penetration factor.  
At large distances the phase   $\varphi=  q  r-\eta \log(2qr) +\sigma $, where $ \sigma $ is 
the Coulomb phase shift. 
 
From the  $K$-matrix  one obtains  the scattering amplitude  ${\cal T}$ by regrouping the standing  waves  into  the regular and outgoing  waves.  
With    $C_c= C(\eta )$ and $C_o=C_f=1$, one has 
\begin{equation} 
\label{d1a} 
{\cal T}^c_{i,j}    =  C_i \left[ - (K^c)^{-1} + f \right]^{-1}_{i,j}\  C_j.     
\end{equation}
In this equation we use a diagonal matrix with $f_{c,c} =2\eta q_c h(\eta) + iq_c C^2(\eta) $ and $f_{0,0}=iq_0$. 

In the single channel case, the textbook  relation of $K^c$ to the  scattering amplitude $\cal {T}$  is~\cite{PRESTON}
\begin{eqnarray} 
\label{b7} 
{\cal T}^c  & = & C^2 (\eta )\left[- \frac{1}{K^c} + 2\eta q h(\eta) + iq C^2 (\eta )\right ]^{-1} \nonumber\\
& \equiv & -[q\cot(\delta)-iq ]^{-1}. 
\end{eqnarray} 
The  atomic level shift is obtained from the  "well known" formula.
\begin{equation}
\label{b9} 
C ^2(\eta ) q \cot(\delta) + 2\eta q h(\eta) = \frac{1}{K^c}
\end{equation}
 and the bound state condition $ \cot(\delta) = i$. 
As found by Trueman~\cite{TRUMAN61} 
 \begin{equation}
\label{b10} 
\epsilon _{1s}  = - \frac{4\pi}{ 2 m } 
\phi  ^2_B(0)   a^c\! \left[ 1  + \frac{a^c}{r_B}(2+ 2\gamma)\right]\!, 
\end {equation}
where  $a^c$ is given by the Coulomb  K-matrix at threshold.
Its relation to the  hadronic  scattering length of the present  model  is
\begin{eqnarray} 
 \label{c8} 
a^c  = & a^{h}\! \left[ 1\! -\! \displaystyle\frac{2R}{r_B}\! +\!  a^{h}\Delta_G\right]
 + \displaystyle\frac{2 m \alpha}{R} b^h  ;&\\
\Delta_G= &\displaystyle\frac{2}{r_B}\left[ 1-2\gamma -\log\left(\displaystyle \frac{2R}{r_B}\right) \right] .
\label {eq:deltaG}&
\end{eqnarray}
XXX
This relation leads to the correction  given by  Eq.~(\ref {deltainfinity}).

In the 2-channel case with   $(i,j) = (c,o)$, the  leading order in the level shift   follows by the replacement of   $ a^{c} $ by the  threshold amplitude in the charged channel $  A^c_{c,c}$, which  is obtained 
 from Eq.~(\ref{d1a}). Including terms to order $q_o^2$ 
\begin{equation} 
\label{d1b} 
A^c_{c,c}=  K^c_{c,c} +  i q_o  (K^{c}_{c,o})^2  -  q_o^2  K^{c}_{o,o}(K^c_{c,o})^2.  
\end{equation}  
At the charged threshold the phase space  left in the open neutral channel is described by the momentum $q_o$. 
Eq.~(\ref{b10}) should now be used with a complex   $  A^c_{c,c}$:
\begin{eqnarray}
\epsilon _{1s}- i \Gamma _{1s}/2  =~~~~~~~~\hspace*{2cm}&& \nonumber \\
- \displaystyle\frac{4\pi}{ 2 m } 
\phi  ^2_B(0)  A^c_{c,c}\left[ 1  + \frac{A^c_{c,c}}{r_B} ( 2+2\gamma)\right].&&\label{d4} 
\end{eqnarray}
In the zero range limit,   the hadronic interaction in the charged channel occurs at a momentum  $q_c^2 = 2m \alpha/R $, while that in the neutral channel still  occurs at the momentum $ q_o$, since the atomic binding energy is negligible. 
The K-matrix elements are energy dependent with 
 a low-energy expansion is analogous to that of Eq.~(\ref{rangeexp}).  
\begin{eqnarray} 
K_{c,c} & = & a^h_{c,c} +  b^h_{c,c}\ \frac { 2m \alpha }{R}, \label{d3x} \\
K_{c,o} &= & a^{h}_{c,o} + \frac{1}{2}(q_c^2  + q_o^2) ~ b^h_{c,o},\label {eq:eff range Kco}\\
K_{o,o}& = & a^{h}_{o,o} + q_0^2   b^h_{o,o}.\label {eq:eff range Koo}
\end{eqnarray} 
We  assume isospin invariance for the range parameters $b^h_{c,c}$ and $b^h_{c,o}$ since they only appear in correction terms.
In the internal region  $ r< R $  the standing waves are 
\begin{eqnarray} 
\label{d2} 
u^i_j   \propto  \sin (q_ir)  \delta_{i,j}  + K_{i,j}  \cos(q_j r).   
\end{eqnarray} 
The  continuity of the wave function matrix $ \hat{u}$ and its logarithmic derivative 
$ \hat{u}^{-1}d\hat{u}/dr $ at the  radius of the charged shell  $R$ gives: 
\begin{eqnarray} 
K^c_{c,c} & = & K_{c,c}\left[ 1- \frac{2R}{r_B} + K_{c,c}\ \Delta_G \right]\ ;\label{e10} \\
K^c_{c,o} & = & K_{c,o}\left[ 1- \frac{R}{r_B} + K_{c,c}\ \Delta_G \right]\ ;\label{e11} \\
K^c_{o,o} & =  & K_{o,o}+ K_{o,c}\ \Delta_G K_{c,o}. \label{eq:Kc-Kt}
\end{eqnarray} 
These corrections are implicit in the single channel equations (\ref{deltapoint2}) and (\ref{deltainfinity}).

The extension to  the  ($\gamma,n$) channel is obtained with the substitution  
$ K^c_{i,j} \rightarrow  K^c_{i,j}   + ( i q_f  K^c_{i,f} K^c_{f,j}) /(   1- i q_f  K^c_{f,f} )$ 
to every  matrix element of the two channel system. 
The higher order terms  in the neutral and photon channels  $K^c_{o,o}$ and  
$K^c_{f,f}$ are negligible such  that 
\begin{equation}  
A^c_{c,c} \approx   K^c_{c,c}    
  + i q_o ( K^c_{c,o})^2 
  + i q_f  (K^c_{c,f})^2 .  
\label {eq:3-channelapprox}
\end{equation} 
The corrections  (\ref{e10}) to (\ref{eq:3-channelapprox}) can be introduced  into the Trueman formula~(\ref{d4}) in order to express it in the form of Eq.~(\ref{Truemanformula}).

 The result for an arbitrary charge distribution follows in  complete
 analogy   to the single channel case discussed in Eq. (\ref{Coulpot}) and following. with the substitutions $R\rightarrow 
 \left< r \right>_{em}$ etc.. 
The corrective terms are given by Eq.~(\ref {deltapoint1}) with
the only change that now $ a^h = a^h_{c,c}$ and $ b^h=b^h_{c,c}$. 
The effects from neutral channels  are negligible. 

The total level width $\Gamma_{1s}$ has two components of comparable magnitude corresponding to the decay via the charge exchange and radiative channel, respectively:
 \begin{equation} 
\label{w4e} 
\Gamma_{1s} = \Gamma_{1s}^{\pi^0n} + \Gamma_{1s}^{\gamma n}.
\end{equation}
They can be physically separated using the Panofsky ratio  $P=$ 1.546(9)~\cite{SPU77}. 
Of special interest in the present context is the hadronic charge exchange channel. 
Here the ratio  $b_i^h/a_i^h$ of Eq.~(\ref{deltai}) is $b^h_{co}/a^h_{co}$, since the charge exchange width  depends quadratically on the amplitude (\ref{eq:eff range Kco}) and
\begin{equation}
\label{w4o}
 \Gamma_{1s}^{\pi^0n}\! =\!  \frac{\Gamma _{1s}}{1\! +\! P^{-1}}\! = \!  
\frac{4\pi }{ m } \phi ^2_B(0) q_o\!  \left[ a^{h}_{c,o}\left(1\! +\!  \delta _{\Gamma }\right)\right] ^2. 
\end{equation}
 Here $\delta _{\Gamma }$ is the counterpart of  $\delta _{1s,i}/2$ of Eq.~(\ref{deltai}): 
\begin{equation}
\delta _{\Gamma }\equiv \frac{1}{2}\delta^{<r>} + \delta^{c} + \frac{1}{2}(q_c^2 +q_o^2) \frac{  b^h_{c,o}}{ a^h_{c,o}} +\frac{1}{2}\delta ^{vp}.\label{jd}
\end{equation}
We have corrected the charge exchange amplitude $a^h_{c,o} $ in Eq.~(\ref{w4o})  not only  for the effective interaction energy in the charged channel (gauge term), but also for the  non-atomic $\pi^0$ energy in the open charge exchange channel (\ref{eq:eff range Kco}).   
This is justified, since this correction can be 'tuned' externally, for example by binding the proton into a potential. 
It is thus of a different nature than the non-trivial corrections for the mass splittings.

\section{Numerical results} 

We now apply these results to the $\pi ^-$p atom. We assume for the moment that the correction for the finite range of hadronic interaction only enters via the range parameter $b$. Isospin invariance is assumed  for  hadronic scattering  parameters  appearing in correction terms. 
The e.m. expectation values appearing in Eq.~(\ref {deltapoint1})   follow from the folded 
($\pi ^-$, p) charge distributions corresponding to the observed form 
factors~\cite {ERI88}: 
$ \left<r\right>_{em}=0.95(1)~{\rm fm};~
\left<1 /r\right>_{em}=1.48(1)~~{\rm fm ^{-1}};
~\left<\log (mr)\right>_{em}=-0.687(9)$ with 
$V_C(0)=\left<\alpha /r\right>_{em}=2.13(2)~{\rm MeV}$.
 We use  the empirical values for the range terms $ ~b_{\pi ^-p}=b_{\pi ^+n}=-0.031(9)  ~m_{\pi }^{-3} ;~b_{\pi ^-n}=b_{\pi ^+p}=-0.058(9)~m_{\pi }^{-3}$ 
 or
$b^+=-0.0044(7)~m_{\pi }^{-3}~; b^-=-0.0013(6)~m_{\pi }^{-3}   $ 
from   $\pi $N scattering data~\cite {HOE83}.
 
The correction terms are given in Table~\ref{table:1}.  For the $\pi ^-p$ energy shift they are obtained from Eq.~(\ref{deltapoint1}) by a two step iteration and do not require the  knowledge of $a_{\pi ^- p}$. 
The  width corrections, calculated from Eqs.~(\ref{w4o}) and (\ref{jd}), require that one knows 
the sign of $a_{c,o}^h$.
 We also give the corrections for the $\pi ^+$p Coulomb scattering length $a^c$~\cite {PRESTON}, which is   similar to the $\pi ^-$p case, but for appropriate sign changes in parameters. 
The $a_{\pi^+ \rm p}$ correction terms follow from our determination of 
$a_{c,c}^h$
and $a_{c,o}^h$ assuming them to be isospin invariant. 
\begin{table*}[htb]
\caption{Coulomb corrections in percent  as described in the text. $\delta ^{vp}$ is included in  $\delta _{1s}$ and  $\delta _{\pi ^+p\rightarrow \pi ^+{\rm p}}$.}
\label{table:1}
\newcommand{\m}{\hphantom{$-$}}
\newcommand{\cc}[1]{\multicolumn{1}{c}{#1}}
\renewcommand{\arraystretch}{1.2} 
\begin{tabular}{@{}lccrcr}
\hline
&Extended charge           & \cc{Renormalization} & \cc{Gauge term} & \cc{Vacuum polarization} & \cc{Total} \\
\hline
$\delta_{1s}$      & $-$0.853(8)& 0.701(4) & $-$0.95(29) & 0.48 & $-$0.62(29) \\
$\delta_{\Gamma} $    & $-$0.427(4) &  0.701(4) & 0.50(23) & 0.24 & 1.02(23) \\
$\delta_{\pi ^+p\rightarrow \pi ^+{\rm p}}$   &  $\ \ $0.853(8)&   0.72(5)&  $-$1.71(29)&  0.48&  0.35(29)\\
\hline
\end{tabular}
\end{table*}
The physical $\pi ^{\pm }$n scattering lengths have Coulomb corrections  of less than $0.1\%$
and  can safely be identified with the hadronic
ones  at the present level of precision.

There is little uncertainty in any of the corrections  within our assumptions. 
It comes  mostly from the experimental value of the range term $b_{\pi ^-p}=b^++b^-$. 
Here the   $b^-$ part contributes 50\% to the error  of the energy shift and nearly all to that of the width. 
 From a purely phenomenological standpoint  its theoretical origin is irrelevant.  
However, to leading order it is  simply generated  by the energy dependence of the Weinberg-Tomozawa amplitude on the one hand and by the nucleon Born term of opposite sign on the other one (Eqs. (44-46) in Ref.~\cite {BER97}), consistent with the experimental value. 
In the case of $\delta_{\Gamma} $, the non-atomic correction in Eq.~(\ref{w4o}) for the neutral pion energy is responsible for 60\% of  the 'gauge term'.

 The  low-energy expansion for the K-matrix depends symmetrically on the initial and final momenta as $(q^2_i+q^2_j)/2$. 
For the  terms  proportional to $b^-$  this is  explicitly the case in the non-relativistic limit when the initial and final 
pion are separately on the mass shell~\cite {BER97}. 
The situation is similar for  the isoscalar effective range term $b^+$.  
The dominant contribution is in this case proportional to the scalar form factor $\sigma (t)$ (see, e.g, Eq.~(10.1) and following in
 Ref.~\cite {GAS89}). 
The corresponding  nucleon Born terms have the same structure. 

We have therefore a good quantitative picture of the precision to which the hadronic scattering length can be extracted from the
 strong energy shift in the limit of a short ranged energy dependent hadronic interaction.  
To fully exploit the present  experimental information content,   the theoretical corrections must controlled at least to  0.5\%.  
This has been achieved even using the errors of  the phenomenological parameters. 
The true theoretical  precision in our approach is far higher.  
The model with a spherical shell  charge distribution suggests that the results are robust with small modifications as long as 
the interaction range is inside a characteristic charge scale.

The correction terms of Table 1 applied to the experimental value of the pionic atom energy shift and width of Eq.~(\ref {epsilonexp})
 give the following hadronic scattering lengths
\begin{eqnarray}  
a^h_{\pi ^- p}~~~~~~ &  =\  a^h_{c,c}\  = & 0.0870(5) ~ m_{\pi }^{-1}~; \label {scattlength}\\
a^h_{\pi ^- p\rightarrow \pi ^0n }& =\  a^h_{c,o}\  = & -0.125(4)~ m_{\pi }^{-1}~.\label {acex}
\end{eqnarray}

Here the masses are the physical ones and the $\pi ^0$n and $\gamma $n decay channels are open.
The result~(\ref{scattlength}) is 1.5\% smaller and outside the quoted uncertainty of the value $0.0883(8)$ deduced 
in Ref.~\cite{SCH01} which is based on the Sigg analysis~\cite{SIGG96}. We have made no correction  in Eq.~(\ref{scattlength})  
for the 
 e.~m. terms outside our Coulomb potential approach, in particular for the 
 effect of the $\gamma n $ and the
 $\gamma \Delta $  intermediate   states in the direct  and crossed  channels. This will be discussed in  the next  section. 
Our value (\ref{acex}) for $a^h_{\pi ^- p\rightarrow \pi ^0n }$  differs by  2.4\% from that 
deduced in Ref.~\cite{SCH01}.

The  scattering length $ a^h_{\pi ^- p} $  can be analyzed jointly with 
 the $\pi^-D$ scattering length   to give  an isovector scattering length  $(a_{\pi ^-p}-a_{\pi ^-n})/2$. 
  We follow the procedure of Ref.~\cite{ERICSON02} with two minor additions. 
First, the triple scattering  term in the multiple scattering  was only partly included. The full triple scattering term according to Refs.~\cite {BEA03,DOR04} represents a contribution of $+0.0027~m_{\pi }^{-1}$~\cite{DOR04} 
 to the theoretical  $\pi^- D$  scattering length in Ref.~\cite{ERICSON02} in the limit of point interactions.
 Following the procedure in
 Ref.~\cite{ERICSON02} we reduce the  overall factor $<1/r^2>_D=0.314\,(25)\, {\rm fm }^{-2}$ in the triple scattering term by a form factor to give $<f(r)^2/r^2>_D=0.238\,(24)\, {\rm fm }^{-2}$. The  theoretical 
 $\pi^- $D 
scattering length in Ref.~\cite{ERICSON02} should therefore be  increased by $+0.0019\,(2)~m_{\pi }^{-1}$.
Second, the Fermi motion ('boost') correction now includes not only the dominant contribution from p-wave $\pi $N scattering, but also a smaller one from the energy dependence of s-wave isoscalar amplitude at threshold as first found in a chiral approach~\cite{BEA03}. In fact, a more accurate description of this  correction term is obtained using the momentum expansion of the forward scattering amplitude near threshold~\cite{ERI04}. 
The p-wave coefficient
 $c_0$ in Ref.~\cite{ERICSON02}, Eq.~(14), should then be replaced  by  $b^++c_0 $ with $b^+=(b_{\pi ^-p}+b_{\pi ^-n})/2$.  The Fermi motion  correction to the $\pi ^- $D scattering length changes then from
$0.0061\,(7)~m_{\pi }^{-1}$ to $0.0047\,(6)~m_{\pi }^{-1}$. The net contribution of these two changes 
to the determination of the isovector scattering length is only about 0.2\% of its numerical value.
This is  small in comparison to the systematic theoretical uncertainty  quoted in Table IV of Ref.~\cite{ERICSON02}. The individual contributions of both changes are also  within this uncertainty.

   Using  $a_{\pi ^-p}$ from  Eq.~(\ref{scattlength}) in conjunction  with the procedure of Ref.~\cite{ERICSON02} Eq.(B8), including  the corrections just described  leads to the improved isovector combination value  
 $(a_{\pi ^-n}-a_{\pi ^-p})/\sqrt{2}$ $=-0.125 \,(1)~m_{\pi } ^{-1}$. It can  therefore
be directly compared to the corresponding  value $ -0.125\,(4)~ m_{\pi }^{-1}$ deduced in 
Eq.~(\ref{acex})  from the charge exchange width
 $\Gamma _{1s}^{\pi^0n}$. 
 Note that when the deuteron data are used to extract the isovector scattering length, the further 
e.~m. contributions cancel in the limit of charge symmetry~\cite{ERICSON02}. Such is the situation for the leading  dispersive contributions from processes such as the  
 $\gamma N(\gamma \Delta )$  channels and their crossed terms. 
 Thus, there is no indication of isospin violation in the isovector amplitude at the present level of precision.

\section{Comparison to previous approaches}

In previous analytical approaches using wave functions little attention was given to  the effect of the 
electromagnetic finite size effects and to the issue of the correct energy of the interaction.  
In  a recent discussion, the Coulomb interaction is cut off  entirely at the range of the strong 
interaction~\cite {HOL99}.  
Several authors starting from Trueman~\cite {TRUMAN61,HOL99} consider the influence of a threshold expansion 
for the hadronic amplitude. 
However, they incorrectly identify the momentum $q_c$  with the Bohr momentum, which leads to a very small 
correction of order $\alpha ^2$. 
The correction to the scattering length proportional to $(a^c)^2$ 
is obtained  to leading order 
$\alpha \log \alpha $, but these previous treatments give incorrectly 
 such  terms of order $\alpha $. 
The numerical consequences  of this difference is small. 
                                     
As discussed in the introduction, the numerical approach  using coupled channels~\cite {SIGG96} correctly
 includes the effect of the finite size and vacuum polarization in the wave function correction as well as 
the renormalization term. It is, however, inconsistent with the low-energy expansion.  In addition,  it makes 
 model dependent corrections for isospin violation and radiative decay effects
(see also comments in Refs.~\cite {LYUBO00,GASSER02}). 
 The numerical result for a single channel does not have these problems. It agrees 
with our explicit result using the same input parameters.

 In Refs.~\cite {LYUBO00,GASSER02}, the energy shift $\epsilon_{1s}$ is related to  the scattering length  in pure QCD.
 The   calculations are performed 
  perturbatively to the same 
 order in $\alpha $ as here  
 in the 
framework of  an effective quantum field theory. The correction to Eq.~(3) 
with this scattering length is obtained  using an expansion in 
in powers of $\alpha$ and of the quark masses $(\bar m_d, \bar m_u)$.
The leading and next-to-leading order terms 
 can be given in closed 
form in terms of $\pi N$ scattering lengths~\cite {LYUBO00}. 
The higher order contributions  have been evaluated 
in the framework of ChPT
 in Refs.~\cite {LYUBO00,GASSER02}.
Since our approach is phenomenological and uses the physical masses, 
the e.~m. EFT  corrections for  mass splittings  are implicitly included in our
scattering length. 
 One cannot therefore match the expansions in detail to each
other. In spite of that it is possible to make some observations.
First, the next-to-leading order non-analytic term  $\alpha\log \alpha$  
 proportional to $(a^h)^2$ (our  renormalization term) results quantitatively also in 
the EFT approach as in all previous potential approaches.
This is natural, since it describes long-range physics.   
Inside our model, the  charge distribution acts as a regulator and leads quantitatively 
to additional  terms of order $\alpha $.

As discussed at the end of the previous section, our Coulomb scattering length $A_{c,c}^c$ in 
Eq.~(29) contains additional non-potential e.~m. contributions, which are an integral
 part of our phenomenological scattering length.  Such contributions have been calculated in
a chiral quark model with quark wave functions~\cite{LYU01}; the authors obtain expressions
 for the EFT empirical constants 
$f_{1,2,3}$ in an approximation corresponding to $\gamma N$ and $\gamma \Delta  $ appearing
 in intermediate states.  The lack of knowledge of  $f_{1}$ is the largest uncertainty quoted in 
the EFT approach ($\pm 2.9\%$)~\cite{LYUBO00,GASSER02}.
 The authors of  Ref.~\cite{LYU01} find that  the constant $f_{1}$ in  their  picture is to 95\% associated 
with the axial baryon form factor.
 The corresponding direct and crossed contribution with $\gamma $n as the intermediate state 
has recently been calculated using soft pion techniques in the heavy baryon limit, 
which gives a  3.4(7)\%  attractive contribution~\cite{ERI03}. Such attraction is also found in the chiral quark model of Ref.~\cite{LYU01}.
 This  follows generically from the predominance of   
intermediate states of energy higher than the threshold one.
 In Ref.~\cite{ERI03} the  leading term  has the symmetry structure corresponding 
to that associated with the ChPT  parameter $f_1$. 
  The   next order term proportional to $m_{\pi }\log m_{\pi }$ has the same coefficient as in  the corresponding term $\delta {\cal T}_3^{em}$ of Ref.~\cite{GASSER02}, which indicates an equivalent physical expansion. 

The main difference of our approach to the e.~m. corrections of the EFT 
approach~\cite {LYUBO00,GASSER02}  is  the wave function and  energy shift corrections which are linear,
 respectively inverse, in the e.~m.  charge radius. In spite of their ready intuitive interpretation,
we have been unable to identify  explicit  counterparts 
in the EFT approach  in which  only even powers $\left< r^{2n}\right>_{em}$
 appear, contrary to our result with a leading linear power.
 This difference occurs already at  the non-relativistic level~\cite {RUS03}.
It is due to a different  handling of the e.~m. charge form~factor, which is seen clearly 
in a configuration space representation. 
We modify the short range part of the Coulomb potential by the form factor and iterate its effect
 on the strong interaction.  
To present order in the expansion the authors of  Refs.~\cite {LYUBO00,GASSER02} consider the
 effect of the form factor  to be equivalent
 to an 
additional short range effective Lagrangian term proportional to $\left<r^2_{em}\right>$; the
 electromagnetic Lagrangian is otherwise taken 
to be independent of the form factor. This generates an e.~m. contribution $a_{fs}$ to the scattering 
length
independent of the strong scattering, which is identical to our result to this order.   
 However, the term in question is generated only by the fact that the charge distribution has a
 range.   Contrary 
to the corresponding situation for a hadronic scattering length   one cannot keep $a_{fs}$ 
(see Eq.~(9))  unchanged in a strict zero range limit, since it would vanish. 
  In addition, the Bohr wave function varies linearly with $r$  near the origin, contrary to the 
 quadratic dependence  for  regular potentials. 
 Both effects make it delicate to iterate a  finite range e. m. interaction 
using an equivalent zero range Lagrangian. 
We conjecture that this difference is only technical. The EFT approach should generate our terms 
in higher orders, since form factors are obtained using  such descriptions.

The $\pi ^-$p scattering length is the dominant contribution in the direct determination 
of the $\pi $NN coupling constant $g_c^2/(4\pi )$ via the GMO sum-rule as given in Eq.~(4) of Ref.~\cite{ERICSON02}. 
Our result (\ref{scattlength}) for the  scattering length, together with 
the reanalysis of the $\pi^-D$ scattering length as described above,  allows an improved  evaluation
 with a new value for  $g_c^2/(4\pi )=14.04\,(17)$,  
  as compared to  $14.11\,(19)$ found in Ref.~\cite {ERICSON02}. 
 As  discussed above,  non-potential terms from the 
  $\gamma $N ($\gamma \Delta $)  channel
 contribute both to $a^h_{\pi ^-p}$ and $a^h_{\pi ^-d}$, but cancel to leading order in 
$m_{\pi }$ in the combination required for the determination~\cite {ERICSON02}.  
Additional contributions  are  within  the systematic  uncertainties.

\section {Conclusions} 

The aim of the present paper was to extract  the 
 phenomenological 
threshold scattering amplitude to high precision and model independently   from the corresponding 
strong interaction energy shift and width in a hadronic atom such as the 
 $\pi ^-$p  atom. Our scattering
 length is differently defined with respect to 
 the QCD one used in Refs.~\cite{LYUBO00,GASSER02}  as discussed above.
We  reach our  goal  within a non-relativistic picture   
by the following key observation. The Coulomb potential of the extended charge distribution is 
perfectly regular at short distances and it is useful to consider it to be an externally applied 
binding potential in addition to the hadronic interaction. This allows us to solve the problem 
exactly for a model with respect to which perturbations can be applied. In this situation the
hadron masses are the physical ones and the extracted hadron amplitude does not assume isospin 
invariance. The regularizing effect of the extended charge is an essential 
feature for the understanding of the correction terms  to the Deser-Trueman relation~(\ref {Truemanformula}). 
It has previously  been included  only in a numerical study using potentials~\cite{SIGG96}.

We show then
 that an accurate relation can only be achieved if three physical effects are properly included.  
First, the relevant wave function at the origin is not the Bohr one, but should correspond to an extended 
charge distribution including vacuum polarization. 
The extended charge distribution is at present important beyond the purely e.~m. energy shift it produces in the atom. 
Second, the correct long-range behavior of the  wave function induces a characteristic change of the wave 
function  near the origin. 
The result  of this feed-back is a quadratic correction to the scattering length proportional to
$\alpha m\ (a^h)^2\log \alpha $ to leading order. 
The leading term is known  from   many investigations.  
Here we obtain  a more accurate result including the  terms of order $\alpha $ with little sensitivity to the model used.
Thirdly,
 the  low-energy expansion of the scattering amplitude leads to a 
characteristic 'gauge' correction which expresses that the scattering 
occurs at an energy typical of that of the extended charge Coulomb 
potential close to the origin~\cite{ERICSON02,Ericson1982}.
This effect is as important as the other corrections.  
Therefore, approaches which do not respect the empirical low-energy expansion  
cannot extract accurate values for scattering lengths from atomic data. 

The present investigation assumes  that the charge distribution produces a Coulomb potential that varies little 
over the range of the strong interaction.
The results of our model in Eq.~(\ref{deltapoint2}) and following suggest that the results are nearly unchanged as 
long as the hadronic range is  within that of the scale set by  the charge distribution.
   Although we have not yet investigated these aspects, the results appear to be robust with compensations between contributions~\cite{ERI04}. 
The sign and magnitude of these corrections are  unlikely to change.
Our  considerations  also apply to other hadronic systems, 
in particular to the $\pi ^+\pi ^-$ atom as will be discussed in a more 
detailed version~\cite{ERI04}.

\noindent \textbf{Acknowledgments.}\\
We  thank Drs.  J. Gasser, A.~Ivanov, V.~Lyubovitskij, E.~Oset, M.~Pavan and A.~Rusetsky,   for  valuable exchange of information.
Drs. A.~Badertscher and G.~C.~Oades kindly let us use their analysis program from Ref.~\cite {SIGG96}.  
B.~L. and S.~W. are grateful for the  hospitality and financial support of the TSL Laboratory, where  this work was partly done.  
S.~W. was supported by KBN grant 5 P 038 04521.
B.~L. acknowledges a grant from the Swedish Royal Academy of Science.

 \end{document}